\documentclass[journal]{IEEEtran}
\usepackage{amsmath,amsfonts}
\usepackage{algorithmic}
\usepackage{algorithm}
\usepackage{array}
\usepackage[caption=false,font=normalsize,labelfont=sf,textfont=sf]{subfig}
\usepackage{textcomp}
\usepackage{stfloats}
\usepackage{url}
\usepackage{verbatim}
\usepackage{graphicx}
\usepackage{cite}
\usepackage{booktabs}
\usepackage{xcolor,colortbl}
\usepackage{multicol}
\usepackage{pbox}
\usepackage{color,soul}
\usepackage{amsmath, bm}
\usepackage{amssymb}
\usepackage{array}
\usepackage{graphicx, nicefrac}
\usepackage{mathrsfs}
\usepackage{dutchcal}
\usepackage{lipsum} 

\newtheorem{theorem}{Theorem}
\newtheorem{lemma}{Lemma}
\newtheorem{proof}{Proof}

\begin{document}

\title{A Novel Stochastic Model for \\IRS-Assisted Communication Systems Based on \\the Sum-Product of Nakagami-$m$ Random Variables}

\author{Hamid Amiriara, Mahtab Mirmohseni,~\IEEEmembership{Senior Member,~IEEE,} Farid Ashtiani,~\IEEEmembership{Senior Member,~IEEE,} and Masoumeh Nasiri-Kenari,~\IEEEmembership{Senior Member,~IEEE}
\thanks{The authors are with the Department of Electrical Engineering, Sharif University of Technology, Tehran, Iran (email: \{hamid.amiriara, mirmohseni, ashtianimt, mnasiri\}@sharif.edu)}
\thanks{This work is based upon research funded by Iran National Science Foundation (INSF) under project No. $4001804$.}}


\IEEEpubid{}

\maketitle

\begin{abstract}
This paper presents exact formulas for the probability distribution function (PDF) and moment generating function (MGF) of the sum-product of statistically independent but not necessarily identically distributed (i.n.i.d.) Nakagami-$m$ random variables (RVs) in terms of Meijer's G-function. Additionally, exact series representations are also derived for the sum of double-Nakagami RVs, providing useful insights on the trade-off between accuracy and computational cost. Simple asymptotic analytical expressions are provided to gain further insight into the derived formula, and the achievable diversity order is obtained. The suggested statistical properties are proved to be a highly useful tool for modeling parallel cascaded Nakagami-$m$ fading channels. The application of these new results is illustrated by deriving exact expressions and simple tight upper bounds for the outage probability (OP) and average symbol error rate (ASER) of several binary and multilevel modulation signals in intelligent reflecting surfaces (IRSs)-assisted communication systems operating over Nakagami-$m$ fading channels. It is demonstrated that the new asymptotic expression is highly accurate and can be extended to encompass a wider range of scenarios. To validate the theoretical frameworks and formulations, Monte-Carlo simulation results are presented. Additionally, supplementary simulations are provided to compare the derived results with two common types of approximations available in the literature, namely the central limit theorem (CLT) and gamma distribution.
\end{abstract}


\section{Introduction}\label{sec1}
\IEEEPARstart{T}{he} performance analysis of digital wireless communication systems over fading channels often involves complex and challenging statistical tasks \cite{ref1}. In the study of intelligent (IRS)-assisted communication systems, accurately determining the distribution of the end-to-end equivalent channel is challenging, particularly when considering the effects of fading channels.
That is, to analyze the fundamental performance limits of IRS-assisted systems, it is necessary to characterize the statistics of the sum of products of independent but not necessarily identical (i.n.i.d.) gamma RVs, or equivalently, squared Nakagami-$m$ random variables (RVs).

\subsection{Related Works and Motivation}
To tackle this open issue, several recent studies have been conducted to investigate IRS-assisted systems, including asymptotic analysis \cite{ref2,ref3,ref4,ref5}, and the use of different approximate methods such as the central limit theorem (CLT)-based approximation \cite{ref6,ref7,ref8,ref9,ref10,ref11,ref12,ref13,ref14}, gamma-based approximation \cite{ref15,ref16,ref17,ref18,ref19,ref20,ref21}, $\mathrm{K_G}$ approximation \cite{ref22,ref23}, Fox's H approximation \cite{ref24}, and mixture of Gaussian (MoG) distribution based approximation \cite{ref_new2}. However, characterizing the performance metrics poses a challenge due to the lack of exact closed-form expressions for the statistical properties. Furthermore, although \cite{ref25} derived exact bit error rate (BER) formulas for the IRS-assisted system with two and three elements, these formulas were expressed as double and quadruple integrals, making them computationally difficult.

The limitations of existing research works also extend to the choice of fading model. Although the Rayleigh fading channel is frequently used in most studies \cite{ref2,ref3,ref4,ref5,ref7,ref9,ref11,ref16,ref22,ref23}, it may not be a suitable model for practical IRS-assisted communication scenarios. This is because IRSs are deployed to exploit line-of-sight (LoS) links between stations, which cannot be accurately modeled by the Rayleigh fading model. Therefore, it becomes necessary to consider Nakagami-$m$ fading channels, which offer a more realistic representation of the practical scenario.
Most of the research works that consider the Nakagami-$m$ fading model employ the CLT and gamma-based approximation methods \cite{ref8,ref10,ref13,ref14,ref18,ref19}. These approximation methods are commonly used for the statistical properties of IRS-assisted systems due to the lack of closed-form expressions. Among these efforts, only references \cite{ref7,ref8,ref10} calculate the related moment generating function (MGF), but even these works rely on the CLT or gamma-based approximations. This highlights the challenges in obtaining exact analytical solutions for the performance metrics in IRS-assisted systems over Nakagami-$m$ fading.

The analysis of diversity order in IRS-assisted communication systems is another problem that existing research works have not sufficiently addressed. While the CLT-based approximation approach has been widely used for this purpose, recent results have shown that this approximation can be inaccurate in certain regimes \cite{ref26}. Furthermore, the gamma-based approximation framework seems inadequate for diversity analysis since it fails to capture the full diversity order \cite{ref24}.

In summary, despite the significant efforts devoted to analyzing the performance of IRS-assisted systems, new analytical methods are necessary to address the limitations of existing works and offer more accurate performance metrics. The MGF plays a crucial role in the calculation of various performance metrics, including the outage probability (OP) and average symbol error rate (ASER). While previous research works have calculated some subset of these metrics, they either lack exact closed-form expressions or rely on approximation-based methods. To the best of our knowledge, this work represents the first attempt to calculate the MGF for the sum-product of i.n.i.d. Nakagami-$m$ RVs, which can be used to derive several performance metrics accurately.

Furthermore, the application of the sum-product of i.n.i.d. Nakagami-$m$ RVs proves highly valuable for modeling parallel cascaded Nakagami-$m$ fading channels, such as parallel multihop relayed communications or other similar applications. This distribution offers a versatile and powerful tool for accurately representing the characteristics of such channels. Therefore, the sum-product of i.n.i.d. Nakagami-$m$ RVs has broader applications in modeling parallel cascaded fading channels, expanding its usefulness beyond IRS-assisted systems.

\subsection{Our Contributions}
To bridge the identified research gap, we propose novel closed-form expressions for the MGF and probability density function (PDF) of the sum-product of i.n.i.d. Nakagami-$m$ RVs. Succinctly, the contributions and novelties of this paper are highlighted as follows:
\begin{itemize}
    \item \textit{Derivation of Exact Formulas:} This paper introduces exact formulas for the PDF and MGF of the sum-product of i.n.i.d. Nakagami-$m$ RVs, leveraging Meijer’s G-function.
    \item \textit{Exact Series Representations:} This work presents exact series representations for the sum of double-Nakagami RVs, highlighting the trade-off between computational cost and accuracy.
    \item \textit{Insightful Asymptotic Expressions:} The research provides simple asymptotic analytical expressions, enriching the understanding of the derived formulas.
    \item \textit{Modeling of Fading Channels:} This study demonstrates that the proposed statistical properties serve as vital tools for modeling parallel cascaded Nakagami-$m$ fading channels.
    \item \textit{Application to Performance Evaluation of IRS-assisted Systems:} The findings apply to derive exact expressions and tight upper bounds for OP and ASER in IRS-assisted communication systems affected by Nakagami-$m$ fading channels.
    \item \textit{Determination of Diversity Order:} This analysis precisely determines the achievable diversity order for an IRS-assisted system operating over a Nakagami-$m$ multipath fading channel.
    \item \textit{Extending the Analytical Framework:} This research establishes that the novel expressions are not only precise but also adaptable to an extended range of scenarios.
    \item \textit{Validation through Simulations:} Through Monte-Carlo simulations, empirical support for the theoretical frameworks is provided.
    \item \textit{Comparison with Existing Approximations:} Additional simulations enable a comparative analysis between the results and well-established approximations in current literature, namely the CLT and gamma distribution.
\end{itemize}

\subsection{Organization and Notations}
The remaining sections of the paper are structured in the following manner. Section~\ref{sec2} focuses on the statistical properties of the sum-product of i.n.i.d. Nakagami-$m$ RVs. We derive closed-form expressions for the MGF and PDF of these RVs, which serve as fundamental tools for the subsequent analysis. In Section~\ref{sec3}, we present a use case that demonstrates the application of our proposed analytical framework for the performance evaluation of IRS-assisted communication systems. We investigate the exact and tight upper bounds of OP and ASER, as well as analyze the diversity order of IRS-assisted systems over a Nakagami-$m$ fading channel. In Section~\ref{sec4}, we explore the generalization of the derived expressions, including their application to IRS-Assisted MISO communication, where the source is equipped with multiple antennas (Section~\ref{sec4A}), and we extend their scope to encompass scenarios involving direct links between the source and destination in IRS-assisted systems (Section~\ref{sec4B}). Section~\ref{sec5} presents the simulation and numerical results. Finally, we conclude the paper in Section~\ref{sec6} with a summary of our findings.

\textit{Notations:} The main notations utilized in this paper are presented as follows. Scalars, vectors, matrices, and sets are denoted by italic letters (e.g., $a$), bold-face lower-case letters (e.g., $\mathbf{a}$), bold-face upper-case letters (e.g., $\mathbf{A}$), and upper-case calligraphic letters (e.g., $\mathcal{A}$), respectively. For any complex vector $\mathbf{a}$, $\angle \mathbf{a}$ denotes a vector in which each element represents the phase of the corresponding element in $\mathbf{a}$, and $[\mathbf{a}]_n$ denotes the $n$-th element of $\mathbf{a}$. The operators $\mathbb{E}[X]$, $f_X(\cdot)$, and $\mathscr{M}_X(\cdot)$ represent the statistical expectation, PDF, and MGF, respectively, for an RV $X$. The exponential and logarithmic functions are represented by $\exp(\cdot)$ and $\log(\cdot)$, respectively. Additionally, $x!$, $\Gamma(x)$, $\left(x\right)_n$, and $G_{p,q}^{m,n}\left[\left.x\right|_{b_1,\ldots,b_q}^{a_1,\ldots,a_p}\right]$ represent the factorial of $x$, the Gamma function \cite{ref27}, the Pochhammer operator\footnote{The Pochhammer symbol, also known as the shifted factorial, is defined as $\left(x\right)_n=\Gamma\left(x+n\right)/\Gamma\left(x\right)$.} \cite[Eq. (8.310)]{ref27}, and the Meijer's G-function\footnote{The Meijer's G-function is defined by the Mellin-Barnes integral as $G_{p,q}^{m,n}\left[x|_{b_1,\ldots,b_q}^{a_1,\ldots,a_p}\right]=\frac{1}{2\pi i}\int_{L}{\frac{\prod_{j=1}^{m}\Gamma\left(b_j-s\right)\prod_{j=1}^{n}\Gamma\left(1-a_j+s\right)}{\prod_{j=m+1}^{q}\Gamma\left(1-b_j+s\right)\prod_{j=n+1}^{p}\Gamma\left(a_j-s\right)}x^s ds}$. The Meijer's G-function is commonly used in the analysis of complex functions, especially in the fields of applied mathematics and physics. It is also used in wireless communication systems to calculate system performance metrics \cite{ref48}.} \cite[Eq. (9.301)]{ref27}, respectively.
Furthermore, $\binom{n}{k}=\frac{n!}{k!(n-k)!}$ and $\binom{n}{k_0,\ldots,k_M}=\frac{n!}{k_0!\ldots k_M!}$ denote the binomial coefficient and multinomial coefficient, respectively. $\left|\cdot\right|$ denotes the absolute value operator, and $U(x)$ denotes the unit step function, defined as $U\left(x\geq0\right)=1$.
Likewise, the modified Bessel functions of the first and second kind of order $\nu$ are denoted by $\mathcal{I}\nu\left(\cdot\right)$ and $\mathcal{K}\nu\left(\cdot\right)$, respectively \cite[Eq. (9.6.3) and Eq. (9.6.2)]{ref28}. The operators $\mathcal{L}\left\{\cdot\right\}$ and $\mathcal{L}^{-1}\left\{\cdot\right\}$ denote the Laplace transform and inverse Laplace transform, respectively.
$\mathcal{O}\left(\cdot\right)$ is the big-O computational complexity notation.
Finally, $\csc(x)$ returns the cosecant of $x$, which is defined as $\csc\left(x\right) = 1/\sin{(x)}$.

\section{Statistical Properties of the Sum-Product of Nakagami-$m$ Random Variables} \label{sec2}
In this section, we present the results concerning the exact and asymptotic statistical characteristics (particularly the PDF and MGF) of the sum-product of i.n.i.d. Nakagami-$m$ RVs. It should be emphasized that the exact and asymptotic PDF and MGF of the sum-product of the Nakagami distribution, derived in the following theorems, are novel and have not been previously reported in the literature to the best of the authors' knowledge. These theorems cover certain special cases utilized in the literature, including the sum-product of independent and identically distributed (i.i.d.) Nakagami RVs \cite{ref10,ref29}, as well as approximations based on the approaches using the CLT \cite{ref6,ref7,ref8,ref9,ref10,ref11,ref12,ref13,ref14} or gamma distribution \cite{ref15,ref16,ref17,ref18,ref19,ref20,ref21}, etc.

We consider $L\times N$ i.n.i.d. Nakagami-$m$ RVs, denoted by $X_{l,n}$ for $l\in\mathcal{L}\overset{\Delta}{=}\left\{1,\ldots,L\right\}$ and $n\in\mathcal{N}\overset{\Delta}{=}\left\{1,\ldots,N\right\}$. The PDF of $X_{l,n}$ is given by
\begin{equation}\label{eq1}
f_{X_{l,n}}\left(x\right)=\frac{2\Omega_{l,n}^{m_{l,n}}}{\Gamma\left(m_{l,n}\right)}x^{2m_{l,n}-1}\exp{\left(-\Omega_{l,n}x^2\right)}U\left(x\right),
\end{equation}
where $m_{l,n}$ is the shape parameter and $\Omega_{l,n}={m_{l,n}}/{\mathbb{E}\left[X_{l,n}^2\right]}$ is the scale parameter. It is important to highlight that in special cases of $m_{l,n}=1$ and $m_{l,n}=\nicefrac{1}{2}$, the PDF in (1) reduces to the one for Rayleigh and one-sided Gaussian distribution, respectively. Without loss of generality, we can assume that the shape parameters, i.e., $m_{1,n}, m_{2,n}, ..., m_{L,n}$, are arranged in ascending order for all $n\in\mathcal{N}$, where $m_{1,n}$ represents the smallest fading parameter.

Next, we define $H_{L,N}$ as the sum-product of $X_{l,n}$ over $l\in\mathcal{L}$ and $n\in\mathcal{N}$:\footnote{The special cases of our general RVs, i.e., $H_{L,N}$, are addressed in \cite{ref2_new} and \cite{ref3_new} for the scenarios when $N=1$ and $L=1$, respectively.}
\begin{equation}\label{eq2}
H_{L,N}=\sum_{n=1}^{N}\prod_{l=1}^{L}X_{l,n}.
\end{equation}

\begin{theorem}[Moments Generating Function]\label{thm1}
The MGF of $H_{L,N}$ can be presented in closed-form as
\begin{align}\label{eq3}
&\mathscr{M}_{H_{L,N}}\left(s\right)=\frac{\pi^{-N/2}}{\prod_{n=1}^{N}\prod_{l=1}^{L}\Gamma\left(m_{l,n}\right)}\notag\\
&\ \ \ \ \ \ \ \ \ \ \ \times\prod_{n=1}^{N}{G_{L,2}^{2,L}\left[\frac{s^2}{4\prod_{l=1}^{L}\Omega_{l,n}}|_{0,0.5}^{\left(1-m_{1,n}\right),\ldots,\left(1-m_{L,n}\right)}\right]}.
\end{align}		
\end{theorem}

\begin{proof}\label{proofthm1}
The MGF of $H_{L,n}=\prod_{l=1}^{L}X_{l,n}$ for all $n\in\mathcal{N}$, as $\mathscr{M}_{H_{L,n}}\left(s\right)=\mathbb{E}_{H_{L,n}}\left[e^{-H_{L,n}s}\right]$, is given by \cite[Appendix]{ref31},
\begin{align}\label{eq4}
\mathscr{M}_{H_{L,n}}\left(s\right)=&\frac{1/\sqrt\pi}{\prod_{l=1}^{L}\Gamma\left(m_{l,n}\right)}\notag\\
&\times G_{L,2}^{2,L}\left[\frac{s^2}{4\prod_{l=1}^{L}\Omega_{l,n}}|_{0, 0.5}^{\left(1-m_{1,n}\right),\ldots,\left(1-m_{L,n}\right)}\right].
\end{align}		
As $H_{L,N}$ in (\ref{eq2}) is the sum of $N$ i.n.i.d. RVs, i.e., $H_{L,n}$, the MGF of $H_{L,N}$ can be obtained by the multiplication of the MGFs of $H_{L,n}$, for all $n\in\mathcal{N}$, as
\begin{equation}\label{eq5}
\mathscr{M}_{H_{L,N}}\left(s\right)=\prod_{n=1}^{N}{\mathscr{M}_{H_{L,n}}\left(s\right)}.
\end{equation}
By substituting (\ref{eq4}) in (\ref{eq5}), (\ref{eq3}) is derived, and the proof is completed.
\end{proof}	

\textbf{Remark~1:} The expression derived in Theorem~\ref{thm1} illustrates that the primary computational complexity lies in computing the Meijer's G-functions. However, as the system parameters, i.e., $m_{l,n}$ and $\Omega_{l,n}$, for $l\in\mathcal{L}$, $n\in\mathcal{N}$, are constants, we can efficiently obtain results by running this function in MATLAB or a table lookup with stored values of the function. Therefore, the computational complexity resulting from the derived expression is not significant \cite{ref14}.

\begin{lemma}[Probability Density Function]\label{lem1}
Applying the inverse Laplace transform to (\ref{eq5}), we can express the PDF of $H_{L,N}$ as
\begin{equation}\label{eq6}
f_{H_{L,N}}\left(h\right)=\mathcal{L}^{-1}\left\{\prod_{n=1}^{N}{\mathscr{M}_{H_{L,n}}\left(s\right)};h\right\}.
\end{equation}
\end{lemma}

It is evident from (\ref{eq6}) that obtaining a closed-form expression for the PDF of the sum-product of Nakagami-$m$ RVs, i.e., $H_{L,N}$, is not feasible due to the complex Meijer's G-function in its MGF, which makes analytical derivations challenging. To address this challenge, we derive an exact series-form expression for the MGF and PDF for the case when $L=2$, corresponding to the sum of double-Nakagami RVs, i.e., $H_{2,N}$, presented in Theorem~\ref{thm2} and Lemma~\ref{lem2}, respectively.
It is worth noting that for several applications, such as the IRS-assisted system, considering $L=2$ is sufficient. Therefore, in this case, the newly derived MGF expression of $H_{2,N}$, i.e., $\mathscr{M}_{H_{2,N}}(s)$, provides a simpler and more practical solution, allowing for an investigation into the trade-off between the computational complexity and the accuracy performance.

Subsequently, we provide a simple asymptotic description of the MGF and PDF for the sum-product of Nakagami-$m$ RVs, i.e., $H_{L,N}$ , in Theorem~\ref{thm3} and Lemma~\ref{lem3}, respectively, which do not involve either the Meijer's G-function or complex series. These accurate expressions will be used in the following sections to evaluate performance metrics straightforwardly.

\begin{theorem}[MGF of Double-Nakagami RVs]\label{thm2}
For double-Nakagami i.n.i.d. RVs, $L=2$, where all $N$ vectors have the same fading parameter, i.e., $m_{1,1}=\ldots=m_{1,N}=m_1$, $m_{2,1}=\ldots=m_{2,N}=m_2$, $\Omega_{1,1}=\ldots=\Omega_{1,N}=\Omega_1$, and $\Omega_{2,1}=\ldots=\Omega_{2,N}=\Omega_2$, the MGF of $H_{2,N}$ can be obtained in closed-form as
\begin{figure*}[!hb]
\hrulefill
\begin{equation}\label{eq8}
\mathscr{F}\left(x\right)\overset{\Delta}{=}{\mathcal{c}^N} \sum_{k_0+k_1+\ldots+k_I=N}\binom{N}{k_0+k_1+\ldots+k_I}\sum_{n_0=0}^{k_0}\ldots \sum_{n_I=0}^{k_I}\left\{\mathcal{k}x\right\},\tag{8}
\end{equation}
where $\mathcal{c}=\frac{2\pi \csc{\left(\left(m_1-m_2\right)\pi\right)}}{\mathrm{\Gamma}\left(m_1\right)\mathrm{\Gamma}\left(m_2\right)}$, $\mathcal{k}=\prod_{i=0}^{I}\left[-\binom{k_i}{n_i}{\mathcal{h}\left(i+m_1\right)}^{n_i}{\mathcal{h}\left(i+m_2\right)}^{k_i-n_i}\right]$, $\mathcal{h}(\mathcal{m})=\frac{{\Omega_1}^\mathcal{m}{\Omega_2}^\mathcal{m}\mathrm{\Gamma}\left(2\mathcal{m}\right)}{\left(1\right)_{\mathcal{m}-m_1}\left(1\right)_{\mathcal{m}-m_2}}$, and $\mathcal{t}=\sum_{i=0}^{I}{n_i\left(m_1-m_2\right)+k_i(i+m_2)}$.
\end{figure*}
\begin{equation}\label{eq7}
\mathscr{M}_{H_{2,N}}\left(s\right)=\mathscr{F}\left(s^{-2\mathcal{t}}\right),
\end{equation}
where the function $\mathscr{F}(\cdot)$ is defined in (\ref{eq8}), at the bottom of this page. In (\ref{eq8}), $I$ denotes the series order, and the series accuracy enhances with increasing values of $I$.
\end{theorem}

\begin{proof}\label{proofthm2}
 The proof is provided in Appendix~\ref{appA}.
\end{proof}	

\textbf{Remark~2:} Since the series expansion in (\ref{eq8}) is derived from the multinomial theorem, its computational complexity is expressed as $\mathcal{O}\left(N^I (I+1)!\right)$. On one hand, this indicates that as the series order, i.e., $I$, increases, the computational demands grow exponentially. On the other hand, the accuracy of this series is tied to the chosen value of $I$, and as $I$ increases, the accuracy improves (as detailed in Appendix~\ref{appA})\footnote{In practical applications, the series expression is highly reliable when $I$ reaches $4$.}. Consequently, our newly derived series expressions present a trade-off between the accuracy and the computational cost in the performance evaluation, depending on $I$. The outcome is contingent upon the choice of $I$ and offers valuable insights for the practical real-time design and implementation of systems.

\begin{lemma}[PDF of Double-Nakagami RVs]\label{lem2}
The new MGF expression derived in Theorem~\ref{thm2} allows for more feasible and efficient analytical derivations. In particular, by applying the inverse Laplace transform to (\ref{eq7}), we can express the PDF of $H_{2,N}$ without the need for evaluating cumbersome integrals or employing any complicated special function, as

\setcounter{equation}{8} 
\begin{equation}\label{eq9}
f_{H_{2,N}}\left(h\right)=\mathscr{F}\left(\frac{h^{2\mathcal{t}-1}}{\Gamma\left(2\mathcal{t}\right)}\right).
\end{equation}
As noted previously, (\ref{eq3}) involves Meijer's G-function, a rather special function which makes it challenging and inefficient to be exploited analytically. Additionally, the computational complexity of the series representation of the MGF in (\ref{eq7}) can significantly increase depending on the value of $I$. However, the OP or ASER at high SNR relies on the MGF of the SNR as it asymptotically tends to infinity, i.e., $s\rightarrow\infty$ \cite{ref1}, \cite[Appendix D]{ref8}.
\end{lemma}

In line with this, in Theorem~\ref{thm3}, we present a novel asymptotic MGF expression for $H_{L,N}$, which is derived without the need for infinite series or complicated special functions.
\begin{theorem}[Asymptotic MGF]\label{thm3}
As $s\rightarrow\infty$, the asymptotic MGF of $H_{L,N}$ is given by
\begin{equation}\label{eq10}
\mathscr{M}_{H_{L,N}}^\infty\left(s\right)=\mathcal{g}s^{-{2\mathcal{t}}^\prime},
\end{equation}
where $\mathcal{t}^\prime$ is defined as the sum of fading parameters for all $N$ vectors, i.e., $\mathcal{t}^\prime=\sum_{n=1}^{N}m_{1,n}$, and $\mathcal{g}$ is defined as
\begin{equation}\label{eq11}
\mathcal{g}\overset{\Delta}{=}\prod_{n=1}^{N}\left\{\frac{2\left(m_{1,n}\right)_{m_{1,n}}\prod_{l=2}^{L}\left(m_{l,n}\right)_{-m_{1,n}}}{\prod_{l=1}^{L}\Omega_{l,n}^{{-m}_{1,n}}}\right\}.
\end{equation}
\end{theorem}

\begin{figure}[!t]
\centering
\includegraphics[width=3.3in]{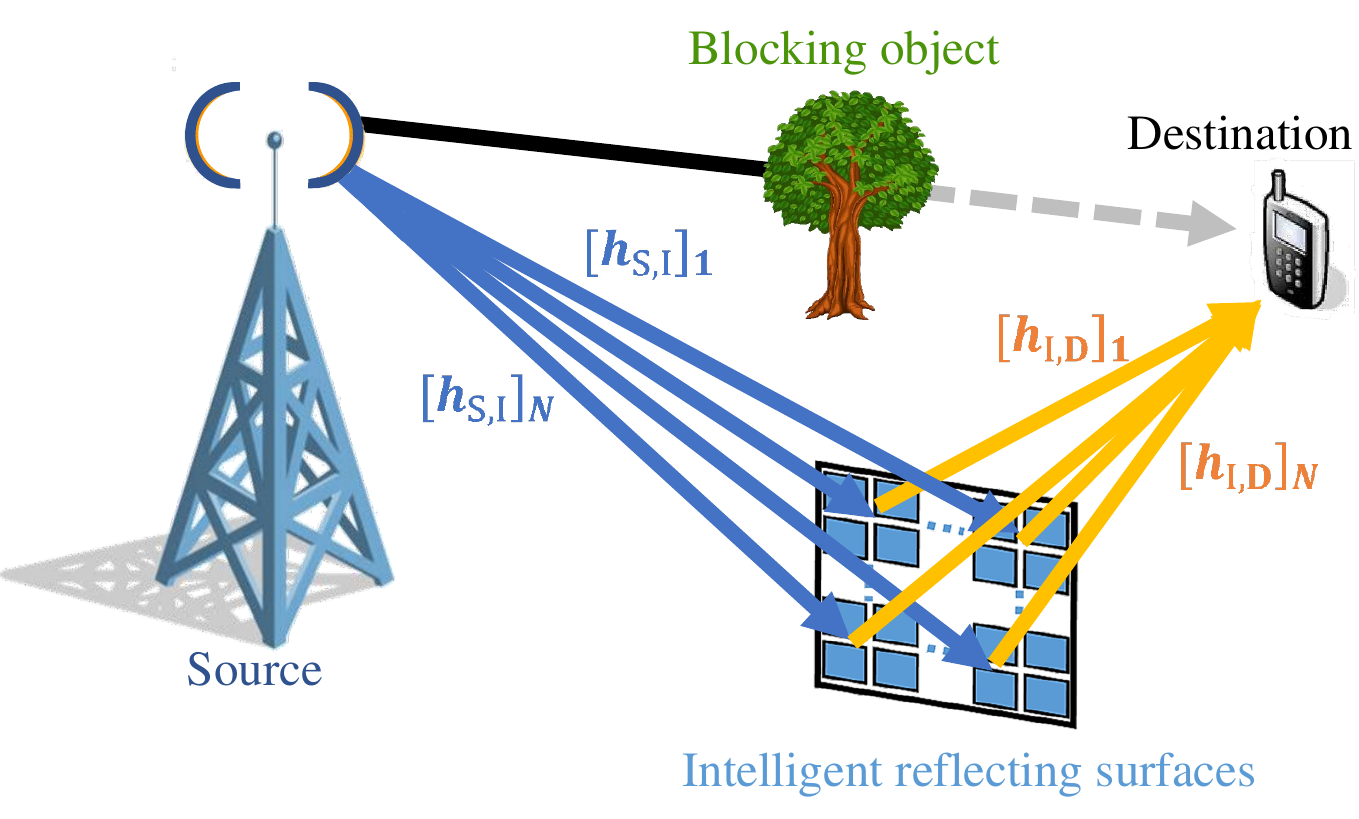}
\caption{IRS-assisted communication system model.}
\label{fig_1}
\end{figure}

\begin{proof}\label{proofthm3}
The proof can be found in Appendix\ref{appB}.
\end{proof}	

\begin{lemma}[Asymptotic PDF]\label{lem3}
By performing the inverse Laplace transform on (\ref{eq10}), the asymptotic PDF of $H_{L,N}$ can be easily obtained:
\begin{equation}\label{eq12}
f_{H_{L,N}}^\infty\left(h\right)=\mathcal{g}\frac{h^{{2\mathcal{t}}^\prime-1}}{\Gamma\left({2\mathcal{t}}^\prime\right)}.
\end{equation}
\end{lemma}

\section{Use Case: Performance Evaluation of Intelligent Reflecting Surfaces}\label{sec3}
In this section, we utilize the previously obtained results to derive both the exact and an upper bound for MGF of the signal-to-noise ratio (SNR) values in IRS-assisted communication systems over Nakagami-$m$ fading environments. As an illustrative application, we present the performance analysis, specifically focusing on the OP and ASER, using these new MGF expressions.

\subsection{System Model}
As depicted in Fig.~\ref{fig_1}, we initially consider a fundamental IRS-assisted Single-Input Single-Output (SISO) communication system comprising a single-antenna source denoted as ``S", a single-antenna destination denoted as ``D", and an IRS denoted as ``I" with $N$ reflecting elements. The transmission process involves the source transmitting a signal, which is then reflected by the IRS before reaching the destination. We make the assumption that the direct link (S-D link) is obstructed by physical obstacles such as trees or buildings \cite{ref32,ref33,ref34,ref35,ref36,ref37}. \footnote{This assumption is particularly realistic, especially in high-frequency scenarios.}

However, in Section~\ref{sec4A}, we expand our analysis to include the case of a multi-antenna source, namely, IRS-assisted MISO communication. In Section~\ref{sec4B}, we further consider the presence of the direct link.

The SNR at the destination receiver can be expressed as \cite{ref37}:
\begin{equation}\label{eq13}
\gamma=\rho\left|\sum_{n=1}^{N}{r_n e^{j\theta_n} \left[\mathbf{h}_{\text{S,I}}\right]_n\left[\mathbf{h}_{\text{I,D}}\right]_n}\right|^2
\end{equation}
where $\rho$ represents the average SNR (ASNR) accounts for the path loss of the end-to-end IRS channel. It is defined as $\rho={l_1l_2p}/{N_0}$, where $p$ denotes the transmitted power, $N_0$ represents the power of the received noise, and $l_1$ and $l_2$ correspond to the path loss for the S-I link and the I-D link, respectively. $\left[\mathbf{h}_{\text{S,I}}\right]_n$ represents the complex channel coefficients between the source and the $n$-th reflecting element of the IRS, and $\left[\mathbf{h}_{\text{I,D}}\right]_n$ denote the complex channel coefficients between the $n$-th reflecting element of the IRS and the destination. It is assumed that $\left|\left[\mathbf{h}_{\text{S,I}}\right]_n\right|$ and $\left|\left[\mathbf{h}_{\text{I,D}}\right]_n\right|$ are RVs that follow the Nakagami-$m$ distribution with shape parameters $m_{\text{S,I}}$ and $m_{\text{I,D}}$, and scale parameters $\Omega_{\text{S,I}}$ and $\Omega_{\text{I,D}}$, respectively. These parameters are assumed to be same for all $N$ elements \cite{ref38,ref39}.

Considering that the IRS is typically deployed in the LoS of the destination, we assume a scenario in which the fading parameter of the S-I link is greater than that of the I-D link, i.e., $m_{\text{S,I}}\geq m_{\text{I,D}}$, without loss of generality.

For the $n$-th reflection element of the IRS, the amplitude attenuation is denoted as $r_n$ and the phase shift as $\theta_n$. To maximize the power of the reflected signal and simplify hardware implementation, we assume that each element perfectly reflects the signal, i.e., $r_n=1$ \cite{ref25,ref32}. Furthermore, we assume that the phase response of each element is locally optimized based on the full channel state information (CSI) available at the IRS, i.e., $\theta_n^\ast=-\angle\left(\left[\mathbf{h}_{\text{S,I}}\right]_n\left[\mathbf{h}_{\text{I,D}}
\right]_n\right)$ \cite{ref25}. Thus, we can rewrite (\ref{eq13}) as
\begin{equation}\label{eq14}
\gamma=\rho\left(\sum_{n=1}^{N}\left|\left[\mathbf{h}_{\text{S,I}}\right]_n\right|\left|\left[\mathbf{h}_{\text{I,D}}\right]_n\right|\right)^2 =\rho\left({H_{2,N}}\right)^2.
\end{equation}

The PDF of the channel coefficient $H_{2,N}$ is the sum of i.n.i.d double-Nakagami RVs. By setting $m_1=m_{\text{S,I}}$, $m_2=m_{\text{I,D}}$, $\Omega_1={\Omega_{\text{S,I}}}$, and $\Omega_2=\Omega_{\text{I,D}}
$ in (\ref{eq9}) and (\ref{eq7}) and applying the transformation method to the two RVs (i.e., $\gamma=\rho\left({H_{2,N}}\right)^2$), the PDF and MGF of $\gamma$ can be obtained as
\begin{equation}\label{eq15}
f_\gamma\left(x\right)=\frac{f_{H_{2,N}}\left(\sqrt{x/\rho}\right)}{2\rho\sqrt{{x}/{\rho}}}=\mathscr{F}\left(\frac{\left({x}/{\rho}\right)^{\mathcal{t}-1}}{2\rho\ \Gamma\left(2\mathcal{t}\right)}\right),
\end{equation}
and
\begin{equation}\label{eq16}
\mathscr{M}_\gamma\left(s\right)=\mathcal{L}\left\{f_\gamma\left(x\right);s\right\}=\mathscr{F}\left(\frac{\left(\rho s\right)^{-\mathcal{t}}}{2\left(\mathcal{t}\right)_\mathcal{t}}\right),
\end{equation}
respectively. However, (\ref{eq16}) involves the function $\mathscr{F}(\cdot)$, which is defined in (\ref{eq8}) and contains a cumbersome series. This makes (\ref{eq16}) challenging to employ for analytical derivations. Thus, we derive an upper bound on the MGF using (\ref{eq10}) as
\begin{align}\label{eq17}
\mathscr{M}_\gamma^\infty\left(s\right)=&\left(\frac{2\left(m_{\text{I,D}}\right)_{m_{\text{I,D}}
}\left(m_{\text{S,I}}\right)_{-m_{\text{I,D}}}}{{{\Omega_{\text{S,I}}}}^{{-m}_{\text{I,D}}}{\Omega_{\text{I,D}}}^{-m_{\text{I,D}}}}\right)^N\notag\\ &\times \frac{\left(\rho s\right)^{-Nm_{\text{I,D}}}}{{2\left(Nm_{\text{I,D}}\right)}_{Nm_{\text{I,D}}}}.
\end{align}	
The newly derived expression is highly accurate and will be utilized in the following sub-sections to evaluate the simple upper bounds of the ASER and OP, as well as for analyzing the diversity order.
\subsection{Outage Probability}
The OP refers to the probability that the SNR falls below a specified threshold value, $\gamma_{\text{th}}$, i.e., $P_{\text{out}}=P_r\left\{\gamma\le\gamma_{\text{th}}\right\}$, which can be determined by utilizing the MGF of SNR \cite[Ch.1]{ref40} as
\begin{equation}\label{eq18}
P_{\text{out}}=\left.\mathcal{L}^{-1}\left\{\frac{\mathscr{M}_\gamma\left(s\right)}{s}\right\}\right|_{\gamma=\gamma_{\text{th}}}.
\end{equation}
By inserting the exact expression of MGF derived in (\ref{eq16}) into (\ref{eq18}), it is possible to obtain the exact OP as
\begin{align}\label{eq19}
P_{\text{out}}&=\mathscr{F}\left(\frac{\rho^{-\mathcal{t}}}{2\left(\mathcal{t}\right)_\mathcal{t}}\left.\mathcal{L}^{-1}\left\{s^{-\mathcal{t}-1}\right\}\right|_{\gamma=\gamma_{\text{th}}}\right)\notag\\
&=\mathscr{F}\left(\frac{\left({\gamma_{\text{th}}}/{\rho}\right)^\mathcal{t}}{2\left(\mathcal{t}\right)_\mathcal{t}\Gamma(\mathcal{t}+1)}\right).
\end{align}		

To avoid mathematical complexity, we propose substituting the simple asymptotic MGF given by (\ref{eq17}) into (\ref{eq18}). After some manipulations, this allows us to derive a straightforward upper bound of OP for an IRS-assisted system over Nakagami-$m$ as
\begin{align}\label{eq20}
P_{\text{out}}^\infty=&\left(\frac{2\left(m_{\text{I,D}}\right)_{m_{\text{I,D}}
}\left(m_{\text{S,I}}\right)_{-m_{\text{I,D}}}}{{{\Omega_{\text{S,I}}}}^{-m_{\text{I,D}}}{\Omega_{\text{I,D}}
}^{-m_{\text{I,D}}}}\right)^N\notag\\
&\times \frac{\left({\gamma_{\text{th}}}/{\rho}\right)^{Nm_{\text{I,D}}}}{{2\left(Nm_{\text{I,D}}
\right)}_{Nm_{\text{I,D}}}\Gamma(Nm_{\text{I,D}}+1)}.
\end{align}		

\subsection{Average Symbol Error Rate}\label{sec3C}
The ASER is a measure of the probability of incorrect detection of a symbol, and it can be calculated for various modulation schemes such as MPAM, MPSK, and MQAM, using the well-known MGF method \cite{ref1}, as:
\begin{equation}\label{eq21}
P_{\text{e}}=\frac{\alpha}{\pi}\int_{0}^{\frac{\pi}{2}}{\mathscr{M}_\gamma\left(\frac{g}{{sin}^2{\phi}}\right)}d\phi,
\end{equation}
where $\alpha$ and $g$ are given in Table~\ref{tab1} \cite[Table 6.1]{ref41}. By substituting the derived exact MGF expression (\ref{eq16}) and simple asymptotic MGF (\ref{eq17}) into (\ref{eq21}), and using the property that $\int_{0}^{{\pi}/{2}}{\sin^{2n}{\varphi}\ d\varphi}=\frac{\sqrt\pi}{2}\left(n+1\right)_{-\nicefrac{1}{2}}$, we obtain the exact and an upper bound on ASER of an IRS-assisted communication system over Nakagami-$m$ channel as
\begin{equation}\label{eq22}
P_{\text{e}}=\frac{\alpha}{4\sqrt\pi}\ \mathscr{F}\left(\frac{\left(\mathcal{t}+1\right)_{-\nicefrac{1}{2}}}{\left(\mathcal{t}\right)_\mathcal{t}}\left(\rho g\right)^{-\mathcal{t}}\right),
\end{equation}
and
\begin{align}\label{eq23}
P_{\text{e}}^\infty=\frac{\alpha}{4\sqrt\pi}&\left(\frac{2\left(m_{\text{I,D}}
\right)_{m_{\text{I,D}}}\left(m_{\text{S,I}}
\right)_{-m_{\text{I,D}}}}{{{\Omega_{\text{S,I}}}}^{{-m}_{\text{I,D}}}{\Omega_{\text{I,D}}}^{{-m}_{\text{I,D}}
}}\right)^N\notag\\
&\times
\frac{\left(Nm_{\text{I,D}}+1\right)_{-\nicefrac{1}{2}}}{\left(Nm_{\text{I,D}}\right)_{Nm_{\text{I,D}}
}}\left(g\rho\right)^{-Nm_{\text{I,D}}},
\end{align}		
respectively.

\subsection{Achievable Diversity Order}\label{sec3D}
In this sub-section, we analyze the diversity order of the IRS-assisted system operating over a Nakagami-$m$ multipath fading channel. Previous studies have determined the diversity order using approximations and bounds, mostly for Rayleigh fading channels \cite{ref8,ref26,ref42}. A commonly used method to analyze the diversity order of IRS-assisted systems is to utilize CLT-based approximations. However, it should be noted that this approach is accurate only when there is a large number of reconfigurable elements \cite{ref4, ref16, ref39}. Additionally, since the distribution of IRS-assisted channel, i.e., $f_{H_{2,N}}\left(h\right)$ shown in (\ref{eq6}), is generally skewed, CLT-based approach is not adequately accurate for high-SNR analysis. However, it is in the high-SNR regime that the analysis of diversity order is of particular interest \cite{ref26}.

Therefore, we utilize the upper bound for the OP in (\ref{eq20}) or the ASER in (\ref{eq23}) to determine the diversity order of IRS-assisted systems operating over Nakagami-$m$ fading channels. The diversity order is deﬁned by the negative slope of the OP or ASER plotted against the ASNR in a log-log scale \cite{ref44}. Therefore, by utilizing (\ref{eq20}) or (\ref{eq23}), we can calculate the diversity order as
\begin{equation}\label{eq24}
div=\lim_{\rho\rightarrow\infty}-{\frac{\log{\left(P_{\text{out}}\right)}}{\log{\left(\rho\right)}}}=\lim_{\rho\rightarrow\infty}-{\frac{\log{\left(P_{\text{e}}\right)}}{\log{\left(\rho\right)}}}=N{m_{\text{I,D}}}
\end{equation}
Specifically, (\ref{eq24}) shows that the diversity order of IRS-assisted systems in Rayleigh fading, i.e., when $m_{\text{I,D}}=1$, is $N$, but it can be greater than $N$ in less severe fading channels, i.e., when $m_{\text{I,D}}>1$.

\definecolor{Color1}{rgb}{0.9,1,1}
\definecolor{Color2}{rgb}{0.9,.9,.9}
\renewcommand\arraystretch{1.3}
\definecolor{Color1}{rgb}{0.9,1,1}
\definecolor{Color2}{rgb}{0.9,.9,.9}
\renewcommand\arraystretch{1.3}
\begin{table}[t!]
\caption{Average Symbol Error Rate for Different Modulation Types with $\alpha$ and $g$ Values.}
\setlength{\tabcolsep}{3pt}
\begin{center}
\begin{tabular}{l l}
\hline
\rowcolor{Color2}\textbf{Modulation Type}&
{${P_e}\left( {\gamma} \right) = \alpha\  Q\left( {\sqrt {2g{\gamma}} } \right)$}
\\[0.1cm]
\hline
\hline
BFSK&
${P_e} = Q\left( {\sqrt {\gamma} } \right)$
 \\[0.1cm]
\rowcolor{Color1}BPSK&
${P_e} = Q\left( {\sqrt {2\gamma} } \right)$
 \\[0.1cm]
QPSK, 4QAM&
${P_e} \approx 2\ Q\left( {\sqrt {\gamma} } \right)$
 \\[0.1cm]
\rowcolor{Color1}MPAM&
${P_e} \approx \frac{{2\left( {M - 1} \right)}}{M}\ Q\left( {\sqrt {\frac{{6\gamma}}{{{M^2} - 1}}} } \right)$
 \\[0.1cm]
MPSK&
${P_e} \approx 2\ Q\left( {\sqrt {2\gamma}\  {{\sin}}\left({\pi}/{M} \right)} \right)$
 \\[0.1cm]
\rowcolor{Color1}Rectangular MQAM&
${P_e} \approx \frac{{4\left( {\sqrt M - 1} \right)}}{{\sqrt M }}\ Q\left( {\sqrt {\frac{{3\gamma}}{{M - 1}}} } \right)\ \ \ \ \ $
 \\[0.1cm]
Nonrectangular MQAM\ \ \ \ \ &
${P_e} \approx 4\ Q\left( {\sqrt {\frac{{3\gamma}}{{M - 1}}} } \right)$
\\[0.1cm]

\end{tabular}
\label{tab1}
\end{center}
\end{table}

\section{Extending the Analytical Framework: Generalization to Diverse Scenarios}\label{sec4}
In this section, we extend and generalize the analytical framework introduced in Section~\ref{sec3}, addressing various aspects of IRS-assisted communication systems to better understand their performance and capabilities.
\subsection{Performance Analysis of IRS-Assisted MISO Downlink Communication}\label{sec4A}
In this subsection, we consider an MISO system model for an IRS-assisted downlink communication network, where the source is equipped with $M$ antennas while the destination is limited to a single antenna due to size constraints.\footnote{It should be noted that this study can be extended to scenarios involving multiple antennas at the destination (SIMO) as well as those involving multiple antennas at both the source and destination (MIMO).} By optimizing the phase shifts at the IRS, we achieve the maximum SNR, expressed as:
\begin{equation}\label{eq_new1}
\gamma=\rho\left(\sum_{m=1}^{M}\sum_{n=1}^{N}\left|\left[\mathbf{h}_{\text{S,I}}\right]_{n,m}\right|\left|\left[\mathbf{h}_{\text{I,D}}\right]_n\right|\right)^2 =\rho\left({H_{2,NM}}\right)^2.
\end{equation}
By comparing (\ref{eq_new1}) to (\ref{eq14}), it can be deduced that they share an identical structure. The only distinction is that, in place of the problem size $N$ in (\ref{eq14}), (\ref{eq_new1}) uses a problem size of $NM$. Hence, the derivations for the SISO scenario are directly applicable. Consequently, the exact and upper bound of the OP for an IRS-assisted MISO communication system over a Nakagami-$m$ channel can be expressed using equations (\ref{eq19}) and (\ref{eq20}), simply by replacing $N$ with $NM$. Furthermore, obtaining the exact and an upper bound of the ASER involves replacing $N$ with $NM$ in equations (\ref{eq22}) and (\ref{eq23}). Additionally, the MISO system's diversity order can be calculated as $div=NMm_{I,D}$.

\subsection{Performance of Intelligent Reflecting Surfaces with Direct Link}\label{sec4B}
In this subsection, we explore our analysis further by incorporating the system model for IRS-assisted communication. We specifically account for the direct link between S and D. In this scenario, data transmission occurs via both the S-R-D reflective link and the S-D direct link. Let ${h_\text{S,D}}$ denote the direct link's channel.
 By optimizing the phase response of each element based on the CSI, i.e., setting $\theta_n^\ast=\angle h_{\text{S,D}}-\angle\left(\left[\mathbf{h}_{\text{S,I}}\right]_n\left[\mathbf{h}_{\text{I,D}}\right]_n\right)$, $\forall n\in\mathcal{N}$, the SNR can be calculated as $\gamma^\text{D}=\rho\left(\left|{h_\text{S,D}}\right|+H_{2,N}\right)^2=\rho\left({H_{2,N}^\text{D}}\right)^2$.
 Since $\left|h_{\text{S,I}}\right|$ and $H_{2,N}$ are independent, the MGF of the end-to-end channel coefficient, i.e., $H_{2,N}^\text{D}\overset{\Delta}{=}\left|h_{\text{S,D}}\right|+H_{2,N}$, can be obtained by multiplying the MGF of the RV $\left|h_{\text{S,D}}\right|$, denoted as $\mathscr{M}_{\left|{h_\text{S,D}}\right|}\left(s\right)$, with the MGF of the RV $H_{2,N}$, denoted as $\mathscr{M}_{H_{2,N}}\left(s\right)$. The derivations for these can be found in (\ref{eq7}) and (\ref{eq45}) in Appendix~\ref{appC}.
By following the same steps as in (\ref{eq18}) and (\ref{eq23}), we can derive the exact OP and ASER in the presence of the direct link.

To determine the upper bounds of OP and ASER in the context of a direct link, we refer to Theorem \ref{thm4} in Appendix~\ref{appC}. This theorem offers a simplified upper bound for the MGF of Nakagami-$m$ RVs. Using this bound, a simplified upper bound for MGF of $\left|{h_\text{S,D}}\right|$, $ \mathscr{M}_{\left|{h_\text{S,D}}\right|}\left(s\right)$ can be derived as
\begin{equation}\label{eq25}
\mathscr{M}_{\left|{h_\text{S,D}}\right|}^\infty\left(s\right)=\frac{2\left({m_\text{S,D}}\right)_{{m_\text{S,D}}}}{{\Omega_\text{S,D}}^{{-m}_\text{S,D}}}s^{-2{m_\text{S,D}}}.
\end{equation}

On the other hand, since we have characterized $H_{2,N}$ as double-Nakagami RVs, we can obtain the asymptotic MGF of $H_{2,N}$ by setting $L=2$, $m_1=m_{\text{S,I}}$, $m_2=m_{\text{I,D}}$, $\Omega_1={\Omega_{\text{I,D}}}$, and $\Omega_2=\Omega_{\text{S,I}}$ in (\ref{eq10}). Therefor, the upper bound of MGF of $H_{2,N}^\text{D}$, can be obtained by multiplying this by (\ref{eq25}) as follows:
\begin{align}\label{eq26}
\mathscr{M}_{H_{2,N}^\text{D}}^\infty\left(s\right)=&\frac{2\left({m_\text{S,D}}\right)_{{{m}_\text{S,D}}}}{{\Omega_\text{S,D}}^{{-m}_\text{S,D}}}
\left( \frac{2\left(m_{\text{I,D}}\right)_{m_{\text{I,D}}}\left(m_{\text{S,I}}\right)_{{-m}_{\text{I,D}}}}
{{\Omega_{\text{S,I}}}^{-m_{\text{I,D}}}{\Omega_{\text{I,D}}
}^{{-m}_{\text{I,D}}}}\right)^N \notag\\
&\times s^{-2\left({m_\text{S,D}}+{N{m_{\text{I,D}}}}\right)}.
\end{align}	

Using the transformation method between two RVs, i.e., $\gamma^\text{D}=\rho{H_{2,N}^\text{D}}^2$, the MGF of $\gamma^\text{D}$ can be obtained as
\begin{align}\label{eq27}
\mathscr{M}_{\gamma^\text{D}}^\infty\left(s\right)=&\frac{\left({m_\text{S,D}}\right)_{{{m}_\text{S,D}}}}{{\Omega_\text{S,D}}^{{-m}_\text{S,D}}}
\left( \frac{2\left(m_{\text{I,D}}\right)_{m_{\text{I,D}}}\left(m_{\text{S,I}}\right)_{{-m}_{\text{I,D}}}}
{{\Omega_{\text{S,I}}}^{-m_{\text{I,D}}}{\Omega_{\text{I,D}}
}^{{-m}_{\text{I,D}}}}\right)^N \notag\\
&\times \frac{\left(\rho s\right)^{-\left({m_\text{S,D}}+{N{m_{\text{I,D}}}}\right)}}
{{\left({m_\text{S,D}}+{N{m_{\text{I,D}}}}\right)}_{{m_\text{S,D}}+{N{m_{\text{I,D}}}}}}.
\end{align}	

Again, by applying the same steps outlined in (\ref{eq18}) and (\ref{eq23}), we can deduce the upper bound of OP and ASER with the inclusion of the direct link as
\begin{align}\label{eq28}
P_{\text{out}}^\infty=&\frac{\left({m_\text{S,D}}\right)_{{{m}_\text{S,D}}}}{{\Omega_\text{S,D}}^{{-m}_\text{S,D}}}
\left( \frac{2\left(m_{\text{I,D}}\right)_{m_{\text{I,D}}}\left(m_{\text{S,I}}\right)_{{-m}_{\text{I,D}}}}
{{\Omega_{\text{S,I}}}^{-m_{\text{I,D}}}{\Omega_{\text{I,D}}
}^{{-m}_{\text{I,D}}}}\right)^N \notag\\
&\times \frac{\left({\gamma_{th}}/{\rho} \right)^{\left({m_\text{S,D}}+{N{m_{\text{I,D}}}}\right)}}
{{\left({m_\text{S,D}}+{N{m_{\text{I,D}}}}\right)}_{{m_\text{S,D}}+{N{m_{\text{I,D}}}}}
\Gamma\left({m_\text{S,D}}+{N{m_{\text{I,D}}}}+1\right)}.
\end{align}	
and
\begin{align}\label{eq29}
P_{\text{e}}^\infty=&\frac{\alpha}{2\sqrt\pi}\frac{\left({m_\text{S,D}}\right)_{{{m}_\text{S,D}}}}{{\Omega_\text{S,D}}^{{-m}_\text{S,D}}}
\left( \frac{2\left(m_{\text{I,D}}\right)_{m_{\text{I,D}}}\left(m_{\text{S,I}}\right)_{{-m}_{\text{I,D}}}}
{{\Omega_{\text{S,I}}}^{-m_{\text{I,D}}}{\Omega_{\text{I,D}}
}^{{-m}_{\text{I,D}}}}\right)^N \notag\\
\notag\\
&\times\frac{\left({m_{\text{S,D}}}+Nm_{\text{I,D}}+1\right)_{-\nicefrac{1}{2}}}
{{\left({m_\text{S,D}}+{N{m_{\text{I,D}}}}\right)}_{{m_\text{S,D}}+{N{m_{\text{I,D}}}}}\left(g\rho\right)^{\left({m_\text{S,D}}+{N{m_{\text{I,D}}}}\right)}},
\end{align}	
respectively. A comparable approach as in (\ref{eq24}) could be applied to indicate that an IRS-assisted system with a direct link achieves a diversity order of ${m_\text{S,D}}+{N{m_{\text{I,D}}}}$. The details are omitted for brevity.
\section{Numerical Results and Discussion} \label{sec5}
In this section, we present simulation results to validate the accuracy of the analytical expressions derived for the PDF, OP, ASER, and the diversity order of IRS-assisted systems in the previous sections. To this end, we conduct Monte-Carlo simulations with $10,000$ runs for each data point. For comparison purposes, we also include two commonly used approximations available in the literature, namely CLT-based and gamma-based approximations. The channel models assume Nakagami-$m$ fading, where the parameters $m_{\text{S,I}}=2$, $m_{\text{I,D}}=m_{\text{S,D}}=1$, and ${\Omega_{\text{S,I}}}=\Omega_{\text{I,D}}=\Omega_{\text{S,D}=1}$. Furthermore, to emphasize the influence of the diversity order, we choose relatively small values for $N$ in our numerical results, similar to \cite{ref4} and \cite{ref24}.
\subsection{Probability Density Function}
\begin{figure}[!t]
\centering
\includegraphics[width=3.5in]{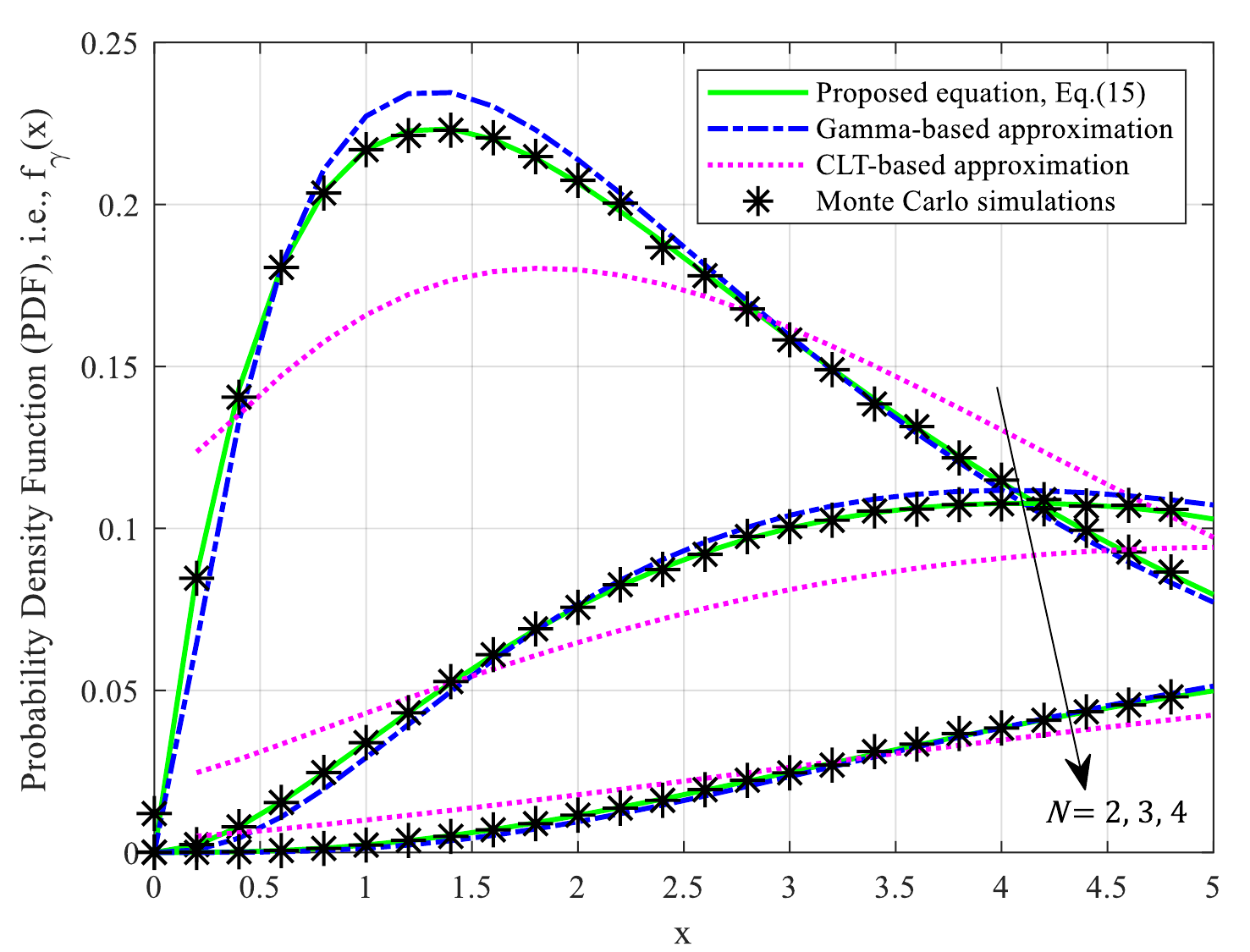}
\caption{Comparison of probability density function of the sum of double-Nakagami RVs for different values of $N$, when the average channel power gain is $\rho=1$.}
\label{fig_2}
\end{figure}

Fig.~\ref{fig_2} illustrates a comparison between the PDF of the ASNR expressed in (15), the CLT-based approximation method \cite{ref37}, gamma-based approximation method \cite{ref21}, and Monte-Carlo simulations. As Fig.~\ref{fig_2} shows, the Monte-Carlo simulations and proposed PDFs (\ref{eq15}) match perfectly, which confirms the accuracy of the derived expression. It can be observed that the proposed PDF is more accurate than both the CLT-based and gamma-based approximations. Additionally, as Fig.~\ref{fig_2} illustrates, increasing the number of reflecting elements, i.e., $N$, enhances the likelihood of achieving higher SNR. This is due to the reduction of multipath propagation, resulting in a more stable and dependable signal with minimal amplitude and phase fluctuations.

\subsection{Average Symbol Error Rate}
\begin{figure}[!t]
\centering
\includegraphics[width=3.5in]{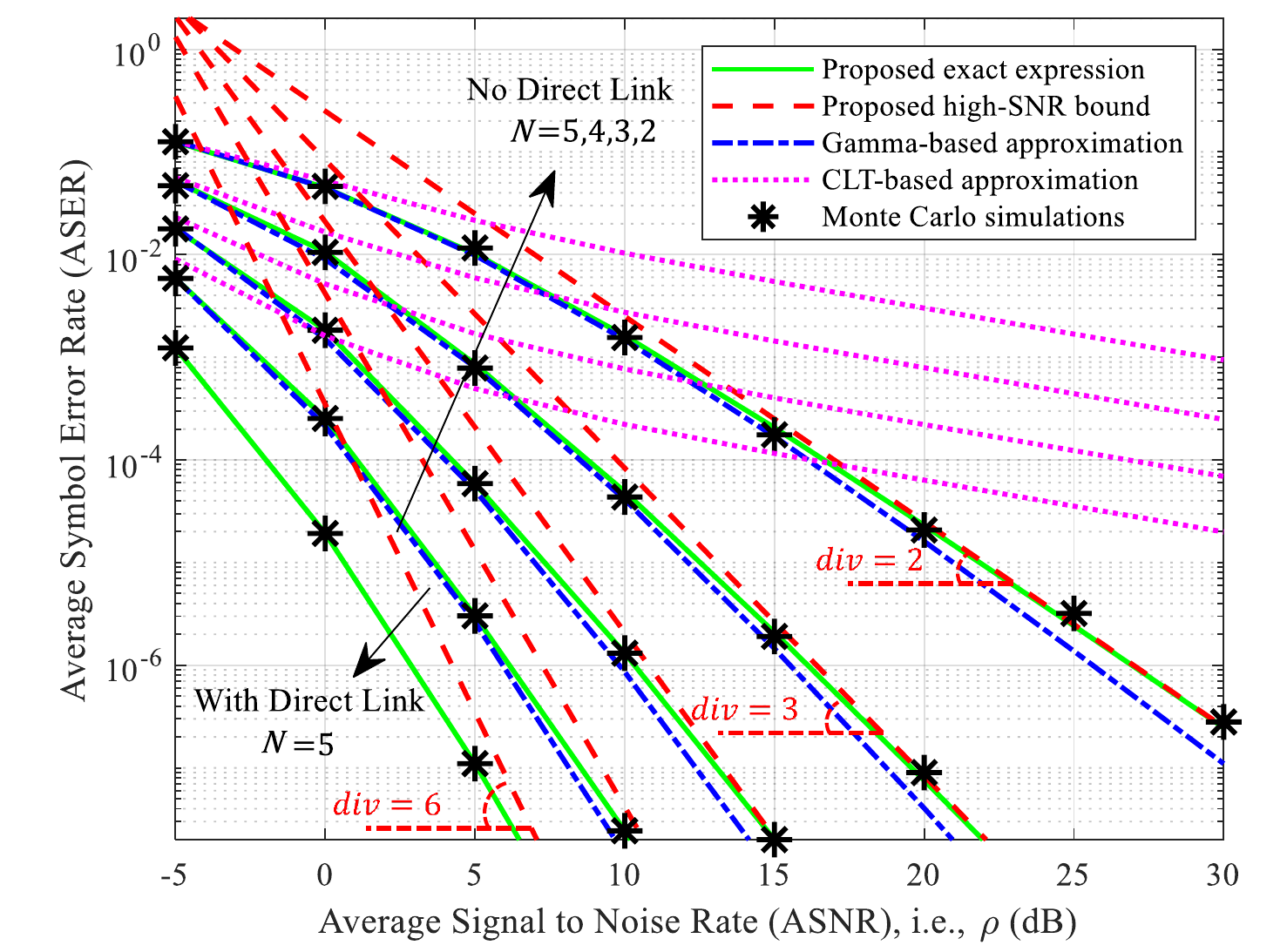}
\caption{An average symbol error rate comparison for different values of $N$ by letting $\alpha=1$ and $g=1$ for BPSK modulation.}
\label{fig_3}
\end{figure}

Figure~\ref{fig_3} illustrates the influence of the ASNR on the ASER for varying numbers of elements in the IRS, $N$. It is noteworthy that we observe a close agreement between the exact expression of the ASER and its corresponding tight upper bound with the results obtained from Monte-Carlo simulations in both scenarios: in the absence and in the presence of the direct link.

Fig.~\ref{fig_3} demonstrates that the CLT-based approximation is reliable at low SNR ranges. However, its accuracy may be compromised due to the skewed distribution of $H_{2,N}$ in (\ref{eq9}). Additionally, the CLT-based approximation has an error floor and becomes constant when $\rho\rightarrow\infty$. Furthermore, as shown in Fig.~\ref{fig_3}, the gamma-based approximation is also unable to provide reliable ASER estimates in the considered setup.

As expected, increasing $N$ results in a decrease in ASER due to the proportional relationship between the ASNR and the number of elements. This means that enlarging the IRS can enhance performance without necessitating additional power consumption. Specifically, the accuracy of the diversity analysis in Section~\ref{sec3D} is confirmed by Fig.~\ref{fig_3}.

Moreover, it is evident from Fig.~\ref{fig_3} that the presence of a direct link, as anticipated, improves the performance.

\subsection{Outage Probability }
\begin{figure}[!t]
\centering
\includegraphics[width=3.5in]{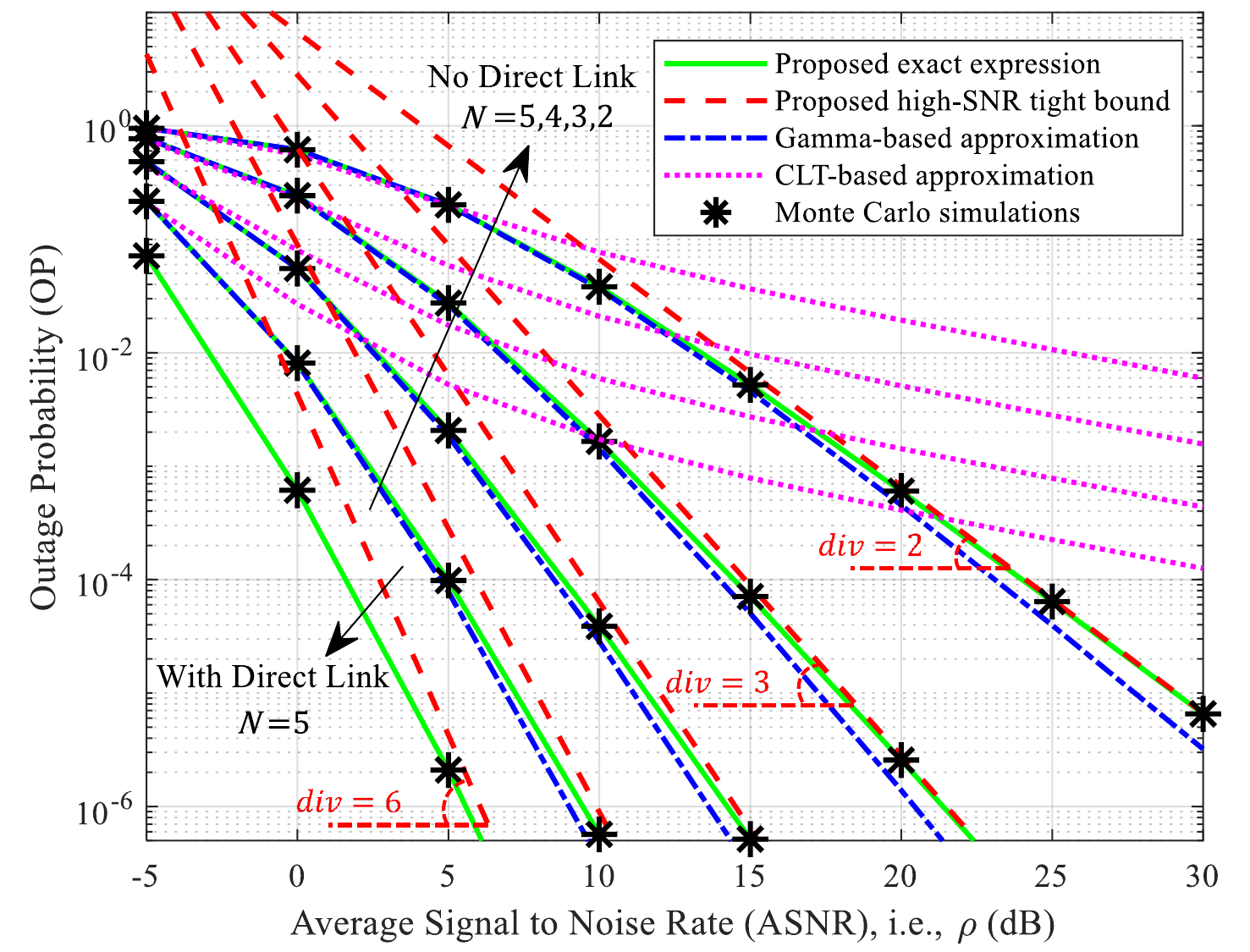}
\caption{Outage probability versus average signal-to-noise ratio with $\gamma_{\text{th}}~=~5 \, \text{dB}$ for different values of $N$.}
\label{fig_4}
\end{figure}

Fig.~\ref{fig_4} depicts OP of an IRS-assisted system as a function of the ASNR for various values of $N$. For both scenarios,in the presence of the direct link and in its absence, the numerical results derived from Monte Carlo simulations are consistent with the analytical findings in Section~\ref{sec3C} and Section~\ref{sec4B}. It is evident from the figure that the asymptotic bound introduced in (\ref{eq20}) serves as a tight upper approximation to the exact OP (\ref{eq19}) for an IRS-assisted system over Nakagami-$m$ fading channels. Furthermore, it is observed that the exact approximation in (\ref{eq19}) and both CLT-based and gamma-based approximations match closely in the low SNR region, but deviate in the moderate and high SNR regions. Consequently, Fig.~\ref{fig_4} confirms that the CLT-based and gamma-based approximations do not yield an accurate estimate of the diversity order in general. As expected, Fig.~\ref{fig_4} also shows that increasing $N$ significantly reduces the required ASNR to achieve a particular OP. For example, to achieve an OP of ${10}^{-4}$, $N=2$ requires an ASNR of $24\, \text{dB}$, while $N=5$ requires only $5\, \text{dB}$.

\section{Conclusion}\label{sec6}
In this paper, we derived an exact formula and tight upper bounds for the PDF and MGF of the sum-product of i.n.i.d. Nakagami-$m$ RVs as well as the accurate formulas for double-Nakagami RVs. These derived statistical properties have proven to be a valuable tool for modeling parallel cascaded Nakagami-$m$ fading channels. Additionally, we introduced a trade-off between computational complexity and accuracy performance. In practical use cases, the proposed MGF serves as a convenient tool for developing a comprehensive analytical framework to analyze the key performance metrics of the IRS-assisted systems operating over Nakagami-$m$ fading channels, including the outage probability (OP), the average symbol error rate (ASER), and the diversity order. Our results were further applied to more general scenarios, such as IRS-assisted MISO communication and IRS-assisted systems with direct links. Moreover, although we focused on the Nakagami-$m$ fading channel, the proposed approach can be extended to other fading channels with similar mathematical structures. Therefore, this work significantly contributes to the understanding of the statistical properties and performance analysis of IRS-assisted systems for future research. That is, the proposed approach can be applied to investigate the performance of the IRS-assisted systems in more complex scenarios, such as multi-user systems, non-linear IRS models, and systems with an imperfect channel state information (CSI). Additionally, the results can be used as a benchmark for the design and optimizing of IRS-assisted systems, which can lead to improved performance and higher spectral efficiency.
\section*{Acknowledgments}
This work is based upon research funded by Iran National Science Foundation (INSF) under project No. $4001804$.

{\appendices
\section{Proof of Theorem~\ref{thm2}}\label{appA}
To find the MGF of $H_{2,N}=\left(H_{2,n}\right)^N$, we first need to express the MGF of the $H_{2,n}$ as a power series.

Note that for $L=2$, the PDF of $H_{2,n}$ can be obtained by applying the inverse Laplace transform to (\ref{eq4}) \cite[Eq. (3.38.1)]{ref45} as,
\begin{equation}\label{eq30}
f_{H_{2,n}}\left(x\right)=\frac{2}{x\ \Gamma\left(m_1\right)\Gamma\left(m_2\right)}\ G_{0,2}^{2,0}\left[x^2\Omega_1\Omega_2|_{m_1,m_2}^-\right],
\end{equation}
By transforming Meijer's G-function into Bessel functions \cite[Eq. (9.34/3)]{ref27}, we show (\ref{eq30}) in terms of modified Bessel function of the second kind,
\begin{equation}\label{eq31}
f_{H_{2,n}}\left(x\right)=\frac{4x^{m_1+m_2-1}b^{m_1+m_2}}{\Gamma\left(m_1\right)\Gamma\left(m_2\right)}\ \mathcal{K}_\nu\left(2bx\right),
\end{equation}
where $\nu=m_1-m_2$ , and $b=\sqrt{\Omega_1\Omega_2}$. The modified Bessel function of the second kind can be represented by modified Bessel function of the first kind as,
\begin{align}\label{eq32}
f_{H_{2,n}}\left(x\right)=&2\pi\csc{\left(\nu\pi\right)}\frac{x^{m_1{+m}_2-1}b^{m_1+m_2}}{\Gamma\left(m_1\right)\Gamma\left(m_2\right)}\notag\\
&\times \left\{\mathcal{I}_{-\nu}\left(2bx\right)-\mathcal{I}_\nu\left(2bx\right)\right\}.
\end{align}	
Using \cite[Eq. (10.25.2)]{ref46}, we can express the modified Bessel function of the first kind as a power series, and (\ref{eq32}) becomes
\begin{align}\label{eq33}
f_{H_{2,n}}\left(x\right)=&\frac{2\pi\csc(\nu\pi)}{\Gamma\left(m_1\right)\Gamma\left(m_2\right)}\notag\\
&\times \sum_{i=0}^{I}\biggl\{\frac{b^{2(i+m_2)}}{i!\ \Gamma\left(i-\nu+1\right)}x^{2(i+m_2)-1}\notag\\
&\ \ \ \ \ \ -\frac{b^{2(i+m_1)}}{i!\ \Gamma\left(i+\nu+1\right)}x^{2(i+m_1)-1}\biggr\},
\end{align}	
where $I$ represents the series order, which is an arbitrary value representing the upper limit of the summation\footnote{While the series expression becomes extremely accurate as the value of $I$ approaches infinity, even a finite $I$ provides satisfactory accuracy in practical applications.}. By applying the Laplace transform to (\ref{eq33}), the MGF of $H_{2,n}$ is given by
\begin{align}\label{eq34}
\mathscr{M}_{H_{2,n}}\left(s\right)&=\frac{2\pi\csc(\nu\pi)}{\Gamma\left(m_1\right)\Gamma\left(m_2\right)}\notag\\
&\times \sum_{i=0}^{I}\biggl\{\frac{b^{2(i+m_2)}\Gamma\left(2(i+m_2)\right)}{i!\ \Gamma\left(i-\nu+1\right)}\left(\frac{1}{s}\right)^{2(i+m_2)}\notag\\
&\ \ -\frac{b^{2(i+m_1)}\Gamma\left(2(i+m_1)\right)}{i!\ \Gamma\left(i+\nu+1\right)}\left(\frac{1}{s}\right)^{2(i+m_1)}\biggr\},
\end{align}	
Since all $N$ RVs possess the same parameters $m_1$, $m_2$, $\Omega_1$, and $\Omega_2$, the MGF of $H_{2,N}$ is given by
\begin{equation}\label{eq35}
\mathscr{M}_{H_{2,N}}\left(s\right)=\left(\mathscr{M}_{H_{2,n}}\left(s\right)\right)^N.
\end{equation}
After utilizing the Multinomial and Multi-binomial \cite{ref47} expressions in (\ref{eq35}) and performing some algebraic manipulations, the proof is completed, and (\ref{eq8}) is derived.

\section{Proof of Theorem~\ref{thm3}}\label{appB}
To prove (\ref{eq10}), first we need to calculate $\mathscr{M}_{H_{L,n}}^\infty\left(s\right)$, since $\mathscr{M}_{H_{L,N}}^\infty\left(s\right)$ can be obtained from (\ref{eq5}). According to the deﬁnition, for the MGF of $H_{L,n}=\prod_{l=1}^{L}{X_{l,n}}$ we have
\begin{align}\label{eq36}
&\mathscr{M}_{H_{L,n}}\left(s\right)=\mathbb{E}_{H_{L,n}}\left[e^{-\left(\prod_{l=1}^{L}X_{l,n}\right)s}\right]=\mathbb{E}_{\prod_{l=2}^{L}X_{l,n}}\biggl[\notag\\
&\ \ \ \ \ \ \ \ \ \ \ \mathbb{E}_{X_{1,n}}\left[e^{-X_{1,n}\left(\prod_{l=2}^{L}X_{l,n}\right)s}|X_{2,n},\ldots,X_{L,n}\right]\biggr]
\end{align}	
where the last step is based on the law of iterated expectations, and $\mathbb{E}_{X_{1,n}}\left[e^{-X_{1,n}\left(\prod_{l=2}^{L}X_{l,n}\right)s}|X_{2,n},\ldots,X_{L,n}\right]$ is the MGF of $X_{1,n}$ scaled by $\prod_{l=2}^{L}X_{l,n}$, i.e., $\mathscr{M}_{X_{1,n}}\left(s^\prime\right)=\mathscr{M}_{X_{1,n}}\left(\left(\prod_{l=2}^{L}X_{l,n}\right)s\right)$. The upper bound of $\mathscr{M}_{X_{1,n}}\left(s^\prime\right)$ can be derived using Theorem~\ref{thm4}, presented in Appendix~\ref{appC}, as
\begin{equation}\label{eq37}
\mathscr{M}_{H_{1,n}}^\infty\left(s^\prime\right)=2\left(m_{1,n}\right)_{m_{1,n}}\  \frac{{\Omega_{1,n}}^{m_{1,n}}}{{s^\prime}^{2m_{1,n}}}.
\end{equation}
Combining (\ref{eq36}) and (\ref{eq37}), we have
\begin{align}\label{eq38}
\mathscr{M}_{H_{L,n}}^\infty\left(s\right)=&2\left(m_{1,n}\right)_{m_{1,n}}\frac{{\Omega_{1,n}}^{m_{1,n}}}{s^{2m_{1,n}}}\notag\\
&\times\mathbb{E}_{\prod_{l=2}^{L}X_{l,n}}\left[\left(\prod_{l=2}^{L}X_{l,n}\right)^{-2m_{1,n}}\right].
\end{align}	
Applying the law of iterated expectations again, the expected value in (\ref{eq38}) can be expressed as
\begin{align}\label{eq39}
&\mathbb{E}_{\prod_{l=2}^{L}X_{l,n}}\left[\left(\prod_{l=2}^{L}X_{l,n}\right)^{-2m_{1,n}}\right]=\mathbb{E}_{\prod_{l=3}^{L}X_{l,n}}\biggl[\notag\\
&\mathbb{E}_{X_{2,n}}\left[X_{2,n}^{-2m_{1,n}}\left(\prod_{l=3}^{L}X_{l,n}\right)^{-2m_{1,n}}|X_{2,n},\ldots,X_{L,n}\right]\biggr].
\end{align}	
On the other hand, since $X_{l,n}$ follows the Nakagami-$m$ distribution with fading parameters $m_{l,n}$ and $\Omega_{l,n}$, the $k$-th moment of $X_{l,n}$ is
\begin{equation}\label{eq40}
\mathbb{E}_{X_{l,n}}\left[{X_{l,n}}^k\right]=\frac{\left(m_{l,n}\right)_{\nicefrac{k}{2}}}{{\Omega_{l,n}}^{\nicefrac{k}{2}}}.
\end{equation}
By inserting (\ref{eq39}) and (\ref{eq40}) into (\ref{eq38}), we have
\begin{align}\label{eq41}
\mathscr{M}_{H_{L,n}}^\infty\left(s\right)=&2\frac{\left(m_{1,n}\right)_{m_{1,n}}}{\Omega_{1,n}^{-m_{1,n}}}\frac{\left(m_{2,n}\right)_{{-m}_{1,n}}}{{\Omega_{2,n}}^{{-m}_{1,n}}}s^{-2m_{1,n}}\notag\\
&\times \mathbb{E}_{\prod_{l=3}^{L}X_{l,n}}\left[\left(\prod_{l=3}^{L}X_{l,n}\right)^{-2m_{1,n}}\right].
\end{align}	
Repeating the same procedure and after some algebraic manipulations, $\mathscr{M}_{H_{L,n}}^\infty\left(s\right)$ can be derived as
\begin{equation}\label{eq42}
\mathscr{M}_{H_{L,n}}^\infty\left(s\right)=2\left(m_{1,n}\right)_{m_{1,n}}\frac{\prod_{l=2}^{L}\left(m_{l,n}\right)_{-m_{1,n}}}{\prod_{l=1}^{L}\Omega_{l,n}^{{-m}_{1,n}}}s^{-2m_{1,n}}.
\end{equation}	
The proof is completed by utilizing (\ref{eq42}) and (\ref{eq5}) to express the MGF of $H_{L,N}$ in closed form as in (\ref{eq10}).
\section{MGF of Nakagami-$m$ RVs}\label{appC}
The exact MGF of a Nakagami-$m$ RV, $X$, with scale parameter $\Omega$ and shape parameter $m$, is defined as:
\begin{equation}\label{eq44}
\mathscr{M}_X\left(s\right)=\mathbb{E}\left[e^{-sX}\right].
\end{equation}
By utilizing the PDF of $X$, given by (\ref{eq1}), (\ref{eq44}) can express as:
\begin{align}\label{eq45}
\mathscr{M}_X\left(s\right) &= 2\ \frac{\Omega^m}{\Gamma\left(m\right)} \int_{0}^{\infty}{x^{2m-1}e^{-sx}e^{-\Omega x^2}dx} \notag\\
&=\frac{\pi^{-1/2}}{\Gamma\left(m\right)}
{G_{1,2}^{2,1}\left[\frac{s^2}{4\Omega}|_{0,0.5}^{1-m}\right]}.
\end{align}
where the final expression in (\ref{eq45}) is a special case of (\ref{eq3}) and is derived from Theorem~\ref{thm1} when $N=1$ and $L=1$.
\begin{theorem}[Upper Bound for MGF of Nakagami-$m$ RVs]\label{thm4}
The upper bound on the MGF of $X$ with scale parameter $\Omega$ and shape parameter $m$ is given by:
\begin{equation}\label{eq43}
\mathscr{M}_X^\infty\left(s\right)=2\left(m\right)_m\frac{\Omega^m}{s^{2m}}.
\end{equation}		
\end{theorem}
\begin{proof}\label{proofthm4}
To establish an upper bound for the MGF of $X$, i.e., (\ref{eq45}), we define two functions: $f\left(x\right)=x^{2m-1}\exp{(-sx)}\exp{\left(-\Omega x^2\right)}$ and $g\left(x\right)=x^{2m-1}\exp{(-sx)}$. It should be noted that as $s\rightarrow\infty$, $f\left(x\right)$ becomes equal to $g\left(x\right)$. It is well-known property that if $f\left(x\right)$ and $g\left(x\right)$ are non-negative functions and $f\left(x\right)\le\ g(x)$ for all $x\geq0$, then $\int_{0}^{\infty}f\left(x\right)dx\le\int_{0}^{\infty}{g(x)dx}$. Therefore, our goal is to demonstrate that $f\left(x\right)\le\ g(x)$ for all $x\geq0$. This can be proven by showing that $\exp{\left(-\Omega x^2\right)}\le\ 1$. Simplifying this inequality yields $\Omega x^2\geq0$, which holds true for all $\Omega\geq\ 0$. Hence, we can establish an upper bound for the MGF of Nakagami-$m$ RVs by utilizing $\int_{0}^{\infty}{g(x)dx}$, as:
\begin{align}\label{eq46}
\mathscr{M}_X\left(s\right)&\le2\frac{\Omega^m}{\Gamma\left(m\right)}\ \int_{0}^{\infty}{x^{2m-1} e^{-sx}dx}\notag\\
&\ =2\frac{\Omega^m}{\Gamma\left(m\right)}\frac{\Gamma\left(2m\right)}{s^{2m}}\overset{\Delta}{=}\mathscr{M}_X^\infty\left(s\right).
\end{align}	
\end{proof}
}

\end{document}